\documentclass[aip,pof]{revtex4-1}

\usepackage{graphicx}
\usepackage{dcolumn}
\usepackage{bm}
\usepackage{psfrag}

\begin{document}

\preprint{APS/123-QED}

\title{Staircase formation in unstably stratified double diffusive finger convection}

\author{A. Rosenthal}

\author{K. L\"udemann}

\author{A. Tilgner}

\affiliation{Institute of Astrophysics and Geophysics, University of G\"ottingen,
Friedrich-Hund-Platz 1, 37077 G\"ottingen, Germany }

\date{\today}

\begin{abstract}
Double diffusive staircases are investigated experimentally in a fluid layer
with a stabilizing temperature gradient and a destabilizing gradient of ion
concentration. Gradients of temperature and ion concentration are maintained
in a steady state within an electrochemical system. Staircases are observed even
if the density stratification is unstable. None of the previously proposed
mechanisms for staircase formation can be recognized in the experiments. Ion
transport through fingers which are part of a staircase is not the same as
transport through fingers extending through the entire cell. Fingers cease to exist if the
diffusive heat transport between neighboring fingers is insufficient.
\end{abstract}

\maketitle

\section{Introduction}

Double diffusive convection \cite{Radko13} is a form of convection in which the density of the
convecting fluid is determined by two different quantities which diffuse at
different rates. The study of double diffusive convection has a long history in
oceanography because the density of ocean water depends on both temperature and
salinity whose diffusivities differ by two orders of magnitude. Double
diffusive convection can give rise to spectacular patterns, such as salt finger
convection in a water layer with stable density stratification, or convection
flows split into vertically stacked layers. These effects motivated numerous
laboratory experiments. Most of these study transient phenomena by setting up
initial temperature and salinity gradients which are then consumed by
convection. Steady states are more difficult to realize. They can be produced
either with a system of membranes and tanks \cite{Krishn03,Krishn09} or with an
electrodeposition cell \cite{Hage10,Kellne14}. The second method is
employed in the experiments presented in this paper.

The electrodeposition cell is filled with a solution of $CuSO_4$ in $H_2SO_4$.
Two copper electrodes placed at the top and bottom of the cell are regulated in
temperature which maintains a temperature gradient within the cell. If a voltage
difference is applied between the upper and lower plates, an electric current
flows through the electrodes and the electrolyte which is carried by copper ions
detached from one electrode and deposited on the other. The copper ion
concentration in the electrolyte plays the role of salinity. However, for copper
ions in an electrolyte to behave like salt in ocean water, the ions should
diffuse and be advected by the flow, but they should not drift in an electric
field due the potential difference between the electrodes. This is the reason
for filling the cell with acid. The highly mobile protons in the acid create
microscopic charged layers next to the electrodes without participating in the
electrochemistry and screen the interior of the cell from any electric field.

This electrochemical system was originally used to study Rayleigh-B\'enard convection
\cite{Goldst90} and later double diffusive finger convection \cite{Hage10} and the
transition from large scale convection rolls to fingers \cite{Kellne14}. The
focus of the new experiments is on double diffusive staircases. These are stacks
of layers filled alternatingly with finger convection or motion organized into
large scale convection rolls. Staircases have been observed both in the oceans
and in laboratory experiments. However, the mechanism by which they form and
hence the conditions necessary for their existence remains controversial. It is
therefore of interest to search for staircases in the electrochemical system.

Layering and staircases in double diffusive systems are studied in various
contexts, for example in chemical engineering and solidification processes
\cite{Kumar18}. Oceanography is the richest field of application, which includes
the study of the properties of staircases once they have formed, as for example
the interaction of internal gravity waves with staircases \cite{Radko20}, and
more fundamentally the study of the mechanisms by 
which staircases form. There are two distinct
cases to be investigated. In the ''diffusive'' case, the salt gradient
stabilizes the fluid whereas the temperature gradient destabilizes it and drives
the motion \cite{Radko16,Brown19,Yang22,Ma22}. The present paper is solely
concerned with the opposite configuration, or the ''finger'' case, with a
stabilizing temperature and a destabilizing salinity gradient. The most
successful theories of staircase formation in this second case are based on a
parametrization of heat and salinity fluxes in terms of local gradients of
temperature and salinity \cite{Radko03,Stellm11,Traxle11}. This approach was
extended to non uniform gradients \cite{Ma21} and received additional numerical
support recently \cite{Yang20}. However, these theories suffer from an
ultraviolet catastrophe in the sense that they yield unbounded growth rates at
high wavenumbers. There is therefore an ongoing effort to improve these theories
\cite{Radko14,Radko19b} and staircase formation remains an open field of
research.

Section \ref{Experiment} reviews the control parameters relevant to the
electrodeposition cell briefly described above and in more detail in
ref. \onlinecite{Hage10}. Section \ref{morpho} qualitatively describes the observed
flows while section \ref{Transitions} discusses various mechanisms which might
possibly induce the transitions between the different types of flows. Section
\ref{Transport} finally deals with the ion transport in the convection flows.

\section{The Experiment}
\label{Experiment}

The material properties characterizing the fluid in the electrodeposition cell are
the kinematic viscosity of the fluid $\nu$, the diffusivities of temperature and
ion concentration, $\kappa$ and $D$, and two expansion coefficients $\alpha$ and
$\beta$ determining variations of density $\rho$ as a function of temperature
$T$ and copper ion concentration $c$ around a reference state with density,
temperature, concentration and pressure $\rho_0$, $T_0$, $c_0$ and $p_0$ via
\begin{equation}
\alpha=-\frac{1}{\rho_0}\left(\frac{\partial \rho}{\partial T}
\right)_{c_0,T_0,p_0}
~~~,~~~
\beta=\frac{1}{\rho_0}\left(\frac{\partial \rho}{\partial c}
\right)_{c_0,T_0,p_0}.
\end{equation}
Both $\alpha$ and $\beta$ are positive. The remaining parameters are the
gravitational acceleration $g$, the cell height $L$, and the temperature and
concentration differences applied across the cell, $\Delta T$ and $\Delta c$,
defined as 
\begin{equation}
\Delta T = T_{\rm{bottom}}-T_{\rm{top}} ~~~,~~~ \Delta c = c_{\rm{top}}-c_{\rm
{bottom}}
\end{equation}
with the subscripts indicating the boundary at which temperature $T$ or
concentration $c$ are evaluated. The concentration difference $\Delta c$ is 
known because the cell is always operated at the limiting current \cite{Hage10,Goldst90}.
In that case, $\Delta c = 2 c_0$, where $c_0$ is the average concentration
of copper in the solution.

Adimensional parameters are more convenient for the comparison with other double
diffusive systems. Four non-dimensional numbers are necessary to specify all 
control parameters. A possible choice for these four parameters are the Prandtl and Schmidt numbers, $Pr$
and $Sc$, defined as
\begin{equation}
Pr=\frac{\nu}{\kappa} ~~~,~~~ Sc=\frac{\nu}{D}.
\end{equation}
together with the thermal and chemical Rayleigh numbers, $Ra_T$
and $Ra_c$, given by
\begin{equation}
Ra_T=\frac{g \alpha \Delta T L^3}{\kappa \nu}  ~~~,~~~ 
Ra_c=\frac{g \beta \Delta c L^3}{D \nu}.
\end{equation}
The material constants $Pr$ and $Sc$ are nearly constant for the measurements
reported here with $Pr \approx 9$ and $Sc \approx 2000$.

With the sign conventions for $\Delta T$ and $\Delta c$, a stabilizing
temperature and a destabilizing concentration gradient are characterized by a
negative $Ra_T$ and a positive $Ra_c$. This combination of signs was realized in
all experiments described in this paper.

The density ratio $\Lambda$ is not independent of the four control parameters
just introduced but is a combination of them which quantifies the ratio of thermal
over chemical buoyancy and which will prove convenient below:
\begin{equation}
\Lambda=\frac{Ra_T}{Ra_c} \frac{Sc}{Pr} = \frac{\alpha \Delta T}{\beta \Delta
c}.
\end{equation}

The observables extracted from the measurements will also be reported in terms
of adimensional numbers. First of all, there is the Sherwood number, which is
directly proportional to the number of ions transported from top to bottom
divided by the purely diffusive current, which means that the Sherwood number is
given by
\begin{equation}
    Sh=\frac{j\,L}{z\,F\,D\,\Delta c}
\end{equation}
if $j$ is the current density, $z$ the valence of the ion ($z=2$ for $Cu^{2+}$),
and $F$ is Faraday's constant. The Sherwood number is
therefore the adimensional analog of the electric current through the cell which
is convenient to measure.

The velocity field was characterized by particle image velocimetry (PIV). A vertical plane near the middle of
the cell was illuminated with a thin laser sheet and observed at right angles with a
camera. In a Cartesian coordinate system in which $z$ points upwards and the
$x$- axis lies horizontally within the plane of illumination, the PIV
measurements yield the velocity components $v_x$ and $v_z$ as a function of
space and time. Let us denote a time average by an overline. 
The Reynolds number $Re$ extracted from these measurements is
defined as
\begin{equation}
Re=\frac{L}{\nu} \left( \frac{1}{A} \int (\overline{v_x^2}+\overline{v_z^2}) dA \right)^{1/2}
\end{equation}
where the integral extends over the entire illuminated plane of surface $A$.
It will be useful for the detection of staircases to define rms amplitudes of
the fluctuations of $v_x$ and $v_z$ as a function of height $z$ as
\begin{equation}
\langle v_x^2 \rangle (z) =
\frac{1}{b} \int_0^b \overline{v_x^2} dx
- \left( \frac{1}{b} \int_0^b \overline{v_x} dx \right)^2
\end{equation}
and
\begin{equation}
\langle v_z^2 \rangle (z) =
\frac{1}{b} \int_0^b \overline{v_z^2} dx
- \left( \frac{1}{b} \int_0^b \overline{v_z} dx \right)^2
\end{equation}
where the integrals extend over the width $b$ of the plane of illumination.

Finally, the PIV measurements also allow the measurement of the finger thickness
$d$ which is expressed in adimensional form as the ratio $d/L$. The finger
thickness is determined by counting the number of changes of sign in the
vertical velocity $v_z$ at any given height $z$ which yields an instantaneous
profile $d(z)$. The time average of this profile is nearly constant over
extended intervals of height within the layers identified below which allows us
to define a finger thickness $d$ for every layer.

\begin{figure}
\includegraphics[width=10cm]{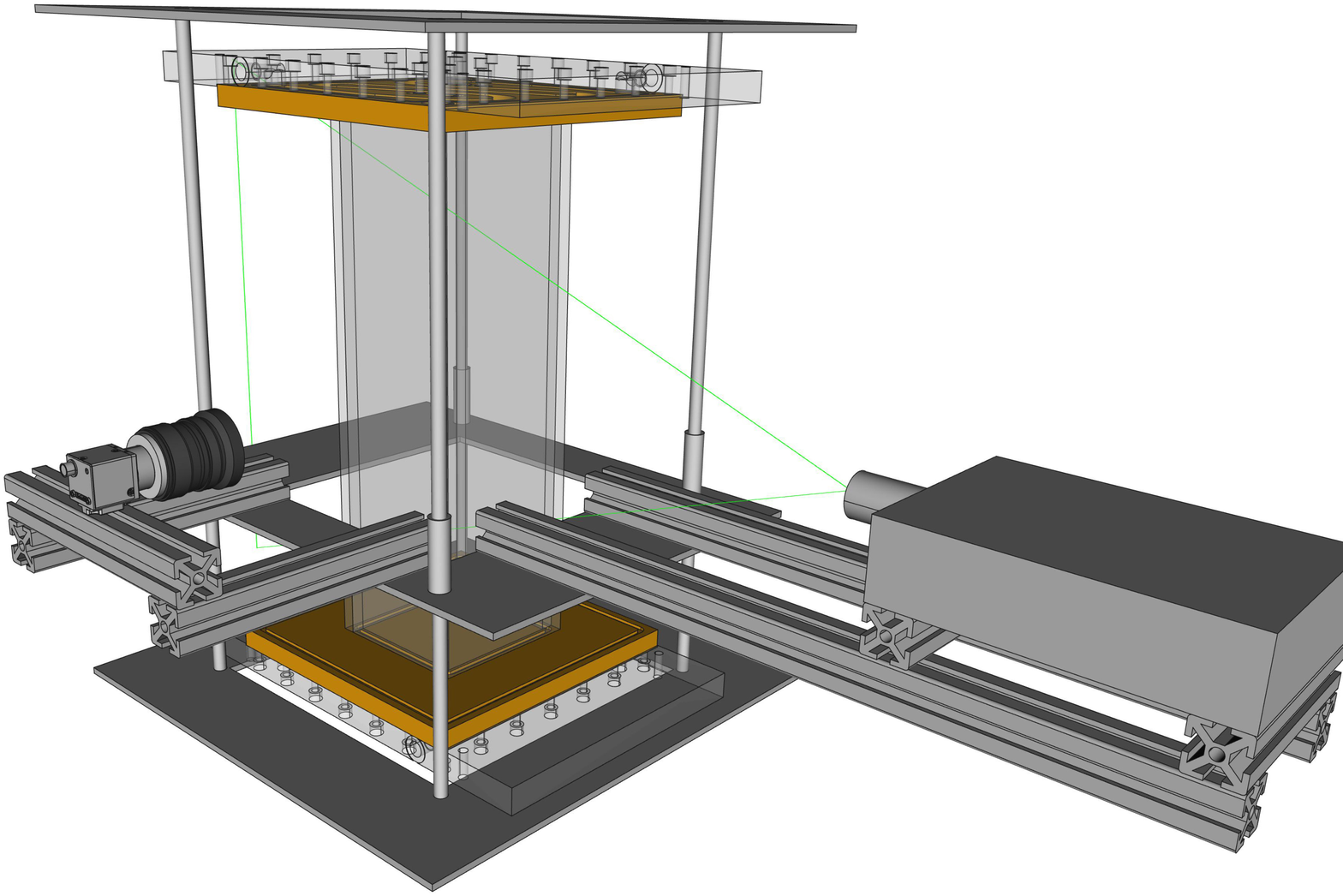}
\caption{(Color online) Schematic view of the apparatus. Top and bottom copper
electrodes are bordered by plexiglass plates for thermal insulation and to
mechanically hold the cell together. Laser and camera can be moved vertically to
observe the flow in the cell at different heights.}
\label{fig:Roboter}
\end{figure}

A schematic view of the apparatus is shown in fig. \ref{fig:Roboter}. Copper
plates at the top and bottom of the cell served as electrodes. A voltage was
applied to the cell with the right polarity to detach copper ions at the top
electrode and to reattach them at the bottom.
Thermally regulated water flowing through channels milled into the copper plates held
the temperature of the electrodes nearly constant and maintained temperature
differences of 1.5-50 $K$ between top and bottom plates. The temperature of the plates
was monitored with thermistors introduced in narrow holes drilled into the
copper. Plate temperature was stable to within $0.1 K$. The cell was filled with
$Cu SO_4$ and 1 molar $H_2SO_4$. 

Ohmic heating in the electrolyte has negligible effect on buoyancy because there
is no electric dissipation in the bulk of the cell which is screened from the
electric field by the protons of the acid. All dissipation occurs next to the temperature
regulated plates in thin layers whose thickness $l$ is on the order of a Debye
length \cite{Probst95}. This length is a few nanometers in the experiments. 
Temperature variation due to
ohmic dissipation is small because heat diffuses rapidly over such a short
length scale. For a quantitative estimate of the temperature perturbation, 
we may model the electric dissipation as heat sources of amplitude $Q$
distributed uniformly within two layers of thickness $l$. If $j$ is the electric
current density and $U$ the total voltage drop across the cell, $Q$ must be
chosen as $Q=j U / (2 l)$ with a factor 2 because there are two layers, one next
to each electrode. To estimate the effect of these heat sources, we solve the
heat diffusion equation $k_T \frac{\partial^2}{\partial z^2}T+Q=0$ with the
boundary conditions that $T=0$ at $z=0$ and $z=l$ and with $k_T$ being the heat
conductivity. This equation is solved by
$T(z)=Qd^2/(2k_T) \cdot z(l-z)/d^2$.
The maximal temperature deviation reached within the layer is therefore
$Q d^2 / (2 k_T)$. It follows that for typical parameters of the experiment 
with $j U < 50 W/m^2$,
$k_T \approx 0.5 W/(K m)$ and $l \approx 10 nm$, the electrolyte is heated by
less than $1 \mu K$ due to ohmic dissipation and that electric dissipation
causes no significant buoyancy in the experiments.

The main difference with the previous experiments on the same system
\cite{Hage10,Kellne14} is the size of the cell. The new experiments used cells
with heights $40cm$ and $80cm$ to reach high Rayleigh numbers. All
cells had a square cross section of $10 cm \times 10 cm$. The field of view of
the camera during PIV measurements could not cover the entire height of the
cell. Camera and laser were therefore mounted on rails so that they could be
moved vertically. Pictures were recorded for several seconds at any given height
level before camera and laser were moved up or down to record a different
section of the cell. This motion was repeated in cycles to avoid any bias that
might be introduced by measuring velocity in one part of the cell at an early
time and in another part of the cell at a late time. Nonetheless, the velocity
measurements were not obtained simultaneously at all heights so that velocity
profiles show discontinuities at the heights separating two sections pictured by
the camera during the measurements. These discontinuities provide us with a
measure of the uncertainty in the velocity measurements due to insufficient time
averaging.

Another feature introduced by taller cells are longer transients because the
turn over time of convection is increased. In addition, it takes longer to form
and equilibrate a staircase than to establish salt fingers spanning the entire
cell, which was the dominant flow structure in the earlier experiments.
The approach to the states containing staircases will be exemplified in fig.
\ref{fig:evolution}. The natural time scale for the transient is the time it
takes fluid within fingers to cross the cell. This transit time was between 15
minutes and 2 hours. Control parameters were typically held constant for one day
which corresponds to 12 to 96 transit times. Long transit times arise if
$|Ra_T|$ is large, which generally results in a single finger layer that reaches
equilibrium after few transit times. The staircases, whose final states are
attained after a more complicated history, occur at smaller $|Ra_T|$ at which
all velocities tend to be larger so that more transit times fit into one day.

\begin{figure}
\includegraphics[width=10cm]{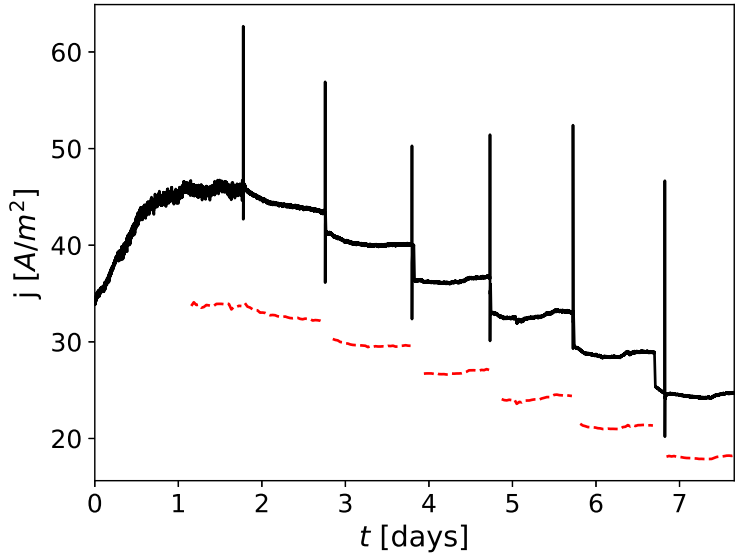}
\caption{(Color online) Current density $j$ as a function of time for a series
of measurements at $Ra_c=1.12 \times 10^{13}$. The current density is computed
as the total current through the cell divided by the cross section of the cell.
The continuous line shows the measured current density, while the dashed line
shows the value of the current density corrected for the roughening of the
electrodes.}
\label{fig:ageing}
\end{figure}

Long transients and data accumulation periods are a problem in the electrochemical system because
the plates age. The extraction and deposition of copper roughens the electrodes
and increases their surface areas which leads to an increased electric current
through the cell. The ageing progresses more rapidly for more concentrated
electrolytes and larger electric currents. However, the roughness of the electrodes
saturates at a certain level. The ratio of the currents through cells equiped
with old and new electrodes never exceeds 1.4.

The effect of ageing was largely avoided in the earlier
experiments by frequently disassembling the cell to polish the electrodes. This
procedure was not practical during the new set of experiments and the electrodes
were left to age for longer before they were refurbished.
One could choose at this stage to declare the cell with roughened electrodes to
be the reference system and to be the main object of study. The precise state of
the boundaries should not matter much if we are interested in a robust effect
like staircase formation which also occurs in the oceans where the boundary
conditions are yet different. But we also want to compare the new results with
the older data obtained with well defined smooth boundaries so that an effort
was made to correct electric current measurements for ageing effects.

A typical series of measurements started by filling the cell equipped with
pristine electrodes with a solution with a certain concentration of cupric
sulfate which fixed the $Ra_c$ for this series. The cell was filled from two
temperature controlled buckets so that the temperature gradient within the cell
immediately had the desired value for the first measurement. $Ra_T$ was subsequently 
varied from one value
to another by modifying the temperature of the copper electrodes while the
voltage was permanently applied. An example of
a time series of the current through the cell from the moment the voltage is
first applied to the next disassembly of the cell is shown in fig.
\ref{fig:ageing}. Short spikes occur in the current when PIV particles are
injected. Each time the temperatures of the boundaries are
modified, there is a rapid variation of the current which corresponds to a
readjustment of the flow in response to the modified control parameters, which
may then be followed by a slow variation of the current mostly due to the
roughening of the electrodes. In order to correct for most of the
ageing effect, intervals of time are marked manually in which the increase of
the current is assumed to be entirely due to the surface modifications of the
boundaries. During any of these intervals, the current changes by a certain
factor. The current measured at all later times is divided by this factor.
This yields a corrected current which is used in the computation of the Sherwood
number. In the example of fig. \ref{fig:ageing}, the current rises already by a
factor of 1.3 between the early times and the first plateau reached
approximately after one day. The flow was a single convection roll during that
period with a transit time of less than 300 seconds. It was therefore assumed
that the variation of current density after $t \approx 1 hr$ was due to
modifications of the electrode and that all surface
roughening occurred during the first day, so that all currents measured thereafter
were divided by this factor before they were introduced into the definition of
the Sherwood number.

\section{Flow morphologies}
\label{morpho}

\begin{figure}
\includegraphics[width=10cm]{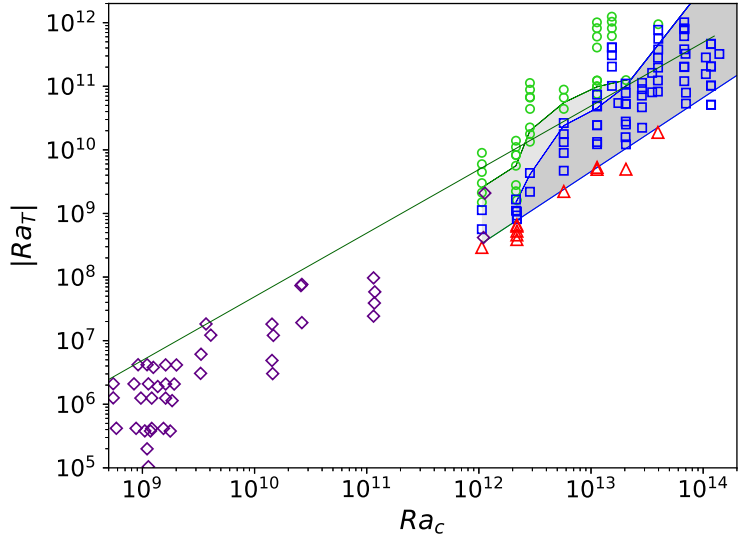}
\caption{(Color online) Overview of the experiments in the $Ra_c,Ra_T$-plane. Diamonds are data
taken from ref. \onlinecite{Hage10}. Circles represent states with fingers
crossing the entire cell from top to bottom which is also the flow morphology
for the diamonds. Squares are staircases defined as
flow states in which the vertical profile of the horizontal rms velocity has a
local maximum. Triangles represent flow states consisting of a convection roll
without any fingers. The dark grey shaded area again indicates the region of
existence of staircases, if a staircase is defined as a flow in which one of the
convecting layers consists of up and down flows with a width comparable with half
the cell width. The lightly shaded area contains all the parameter combinations
for which the Sherwood number depends weakly on $|Ra_T|$ (as discussed in
connection with fig. \ref{fig:Knick}). The straight line indicates $|\Lambda|=1$.}
\label{fig:param_space}
\end{figure}

Fig. \ref{fig:param_space} presents an overview of the combinations of $Ra_T$
and $Ra_c$ sampled during this and a previous study \cite{Hage10}. The simplest
types of flows found in the explored part of the parameter space are double
diffusive fingers crossing the entire cell from top to bottom and convection
rolls reminiscent of Rayleigh-B\'enard convection. Fingers exist both at
$|\Lambda| > 1$ and $|\Lambda| < 1$ even though the density stratification is
unstable in the latter case. There is a well defined transition in going from 
fingers to convection rolls \cite{Kellne14} discussed further in section
\ref{Transitions}. 

Numerous laboratory and field observations suggest that it should also be
possible to generate double diffusive staircases in the electrodeposition cell.
Staircases indeed appear, but only if $Ra_c > 2 \times 10^{12}$. A similar
threshold of $Ra_c \approx 10^{11}$ has been found in numerical simulations
\cite{Yang20} in 3D periodic boxes with $Sc=21$ and $Pr=7$. The staircases in
the experiment consist of a convection roll sandwiched between two finger layers
each in contact with one of the boundaries. Staircases with more layers never developed. 
There are a few examples at the low $|Ra_T|$ end of the region of existence of staircases in fig.
\ref{fig:param_space} in which the staircase forms only two layers, a convection
roll above or below a finger layer.

\begin{figure}
\includegraphics[width=10cm]{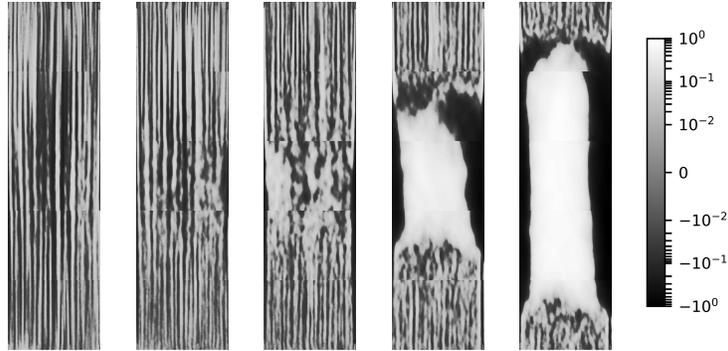}
\caption{Grey scale maps of the vertical velocity normalized by the maximal
absolute value of vertical velocity for $Ra_c=1.12 \times 10^{13}$ and, from left to right,
$|Ra_T|=1.22 \times 10^{11}$, $1.01 \times 10^{11}$, $4.97 \times 10^{10}$,
$2.44 \times 10^{10}$ and $1.23 \times 10^{10}$. Each panel is a time average
over two hours.}
\label{fig:Ra_sequence}
\end{figure}

\begin{figure}
\includegraphics[width=10cm]{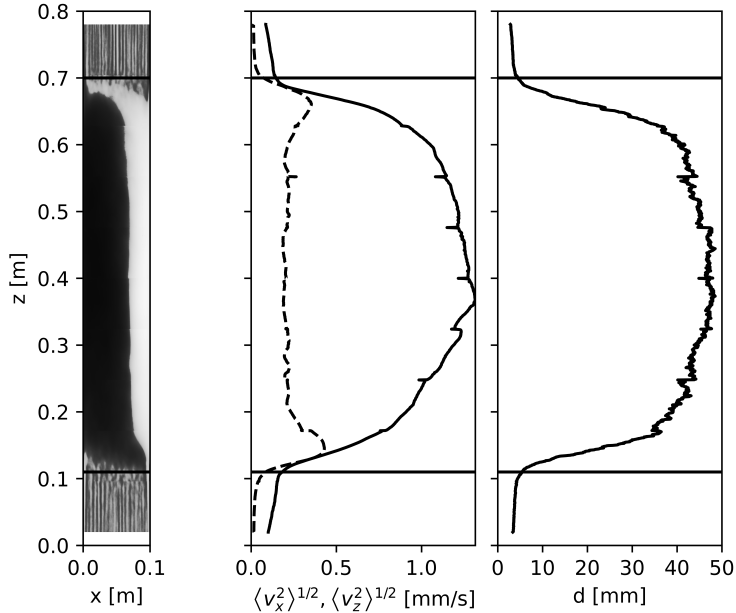}
\caption{Map of the vertical velocity (left panel), variations of rms of
horizontal and vertical velocity with height (central panel, dashed and
continuous lines, respectively), and finger thickness as a function of height
(right panel) for $Ra_c=6.89 \times 10^{13}$ and $|Ra_T|=3.24 \times 10^{11}$. All panels show results obtained
from averaging over 46 hours. The horizontal lines show the heights that were
identified as limits of the upper and lower finger layers according to the
criterion described in the text.}
\label{fig:vh_profile}
\end{figure}

As opposed to the very clear cut transition between flows consisting of a single
convection roll and a single layer of fingers, the transition between the single
finger layer and the staircase is more blurred and passes through intermediate
stages. If $|Ra_T|$ is varied from large to small values at fixed $Ra_c$ with
$Ra_c > 2 \times 10^{12}$ in fig. \ref{fig:param_space} (as shown in fig.
\ref{fig:Ra_sequence} for $Ra_c=1.12 \times 10^{13}$), one first finds straight fingers crossing the
whole cell, but at lower $|Ra_T|$, the fingers thicken in the central region of
the cell. At yet lower $|Ra_T|$, one visually identifies a central finger layer
with thicker fingers than in the adjacent layers above and below. Finally, at
even smaller $|Ra_T|$, the central layer is filled by a convection roll. The
ratio of height over width of the central region is always around 1 or larger, so
that an obvious definition of a convection roll in this context is a flow
consisting of up and downwellings of roughly half the cell width. There is
however some arbitrariness in what one wishes to call a staircase. A plausible
criterion for the edge of a finger layer is the existence of a local maximum in
the rms of horizontal velocity, because the vertical flow within the fingers
needs to connect through horizontal motions at the ends of the fingers. However,
as the example in fig. \ref{fig:vh_profile} shows, the maximum in the rms of
horizontal velocity lies well within the convection roll and marks the edge of
the velocity boundary layer inside the convection roll rather than the edge of
the finger layer. A better indicator for the size of a finger layer therefore is
a kink or sudden change in first derivative of the vertical profile of the rms
of horizontal velocity, paired with plain visual inspection of PIV images. The
height of finger layers $L_f$ determined in this fashion will be used below.

As finger layers in staircases may border a convection roll or a less clearly
classified internal layer, fig. \ref{fig:param_space} shows the region of
existence of staircases according to two different definitions: 
staircases whose central region consist of
a broad up and down flow reminiscent of Rayleigh-B\'enard convection, and
staircases in which a local maximum in the profile of horizontal rms velocity signals the end of a
finger layer. The former region is of course contained within the latter.

No hysteresis could be detected in any of the transitions mentioned above.
$Ra_T$ was varied back and forth on several occasions and the final state was
always independent of the history of the flow.

The most important qualitative result of this section is that staircases can
exist at both $|\Lambda| > 1$ and $|\Lambda| < 1$. Any motion in the bottom
heavy situation $|\Lambda| > 1$ is only possible thanks to double diffusive
effects and requires finger formation somewhere in the volume. A top heavy layer
is naively expected to undergo conventional convection, but this is not always
observed. Both single finger layers as well as staircases may form if $|\Lambda|
< 1$, and more expectedly also if $|\Lambda| > 1$. However, staircases are only
possible if $Ra_c$ exceeds a critical value.

\section{Transitions}
\label{Transitions}

\subsection{Transition from convection rolls to fingers or staircases}

This transition is crossed by increasing $|Ra_T|$ starting from small values at
constant $Ra_c$ in fig. \ref{fig:param_space}. The previous experimental study
\cite{Kellne14} of this transition identified two possible criteria for this
transition. The first criterion was that the condition $|\Lambda|=1/30$ 
delimited the convection rolls from the
finger flows. At the same time, the two types of flows could be predicted
according to the following argument: fingers only form if there is little
exchange of ions between neighboring fingers while heat rapidly diffuses between them.
Heat has to diffuse a distance larger than the finger thickness during the time
it takes fluid to advect vertically from one end of a finger to the other. The
ratio of diffusion length over finger thickness is given by
$\frac{1}{d} \sqrt{\kappa L_f / V}$ where $V$ is the finger velocity. This ratio
was found to be 1 at the transition from convection rolls to fingers if $V$ is
determined from the square root of twice the kinetic energy in the flow.

\begin{figure}
\includegraphics[width=10cm]{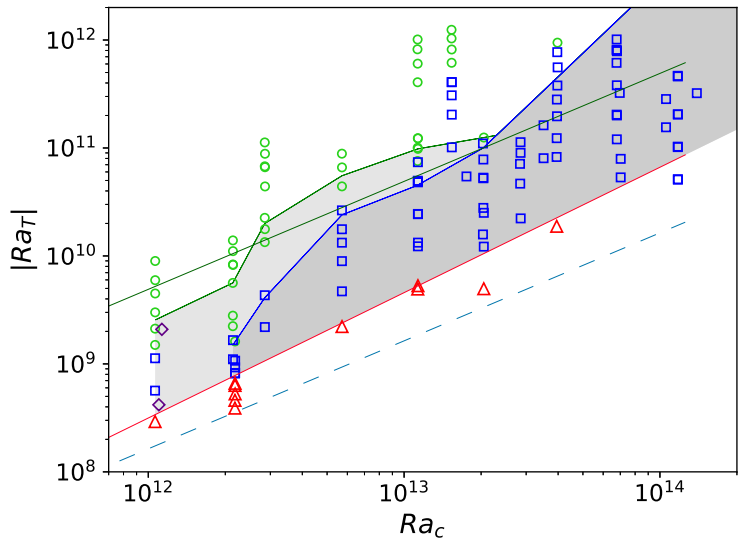}
\caption{(Color online) Excerpt of fig. \ref{fig:param_space} emphasizing the transition from
convection rolls (triangles) to fingers (circles) and staircases (squares)
at large $Ra_c$ and $|Ra_T|$. The dashed line indicates $|\Lambda|=1/30$ and the
lower straight solid line represents $\frac{1}{d} \sqrt{\kappa L_f / V}=1$, whereas the
upper straight solid line is again at $|\Lambda|=1$.}
\label{fig:transition1}
\end{figure}

The data base is extended by several orders of magnitude in $Ra_c$ and $|Ra_T|$
by the new experiments. At large $Ra_c$, the transition is between a convection
roll filling the entire cell to a staircase with short fingers next to one or
both plates so that the finger height $L_f$ may in general be less than the cell
height $L$. As fig. \ref{fig:transition1} shows, the criterion $|\Lambda|=1/30$
is untenable in light of the new data, while 
$\frac{1}{d} \sqrt{\kappa L_f / V}=1$ accurately describes the transition
between convection rolls and fingers.

\subsection{Transition from fingers to staircases}

This transition is crossed by decreasing $|Ra_T|$ starting from large values at
constant $Ra_c$ in fig. \ref{fig:param_space}. We cannot hope for a quantitative
prediction of the transition line of the same quality as in the previous section
because the transition itself is not as sharp. However, the very existence of
staircases at $|\Lambda| < 1$ is noteworthy so that more qualitative
considerations are also appropriate. 

The mechanisms for staircase formation proposed in the past fall into several
categories. There are those which postulate a lateral inhomogeneity
\cite{Merryf00} which likely exists in the oceans but which is not a plausible
ingredient in a controlled laboratory environment. Another class of proposals is based on
double diffusive effects in stably stratified media, such as the instability of
internal gravity waves \cite{Stern69,Holyer81} or a stability criterion based on
a Richardson number \cite{Kunze87}. All these seem to be of little interest here
since we observe staircases also in the unstably stratified case $|\Lambda| <
1$.

A mechanism that turned out to be significant in recent simulations
\cite{Stellm11,Yang20} is an instability which appears \cite{Radko03} if the
flux ratio has a minimum as a function of $|\Lambda|$. In that case, the
difference between heat and salinity fluxes generates an unstably stratified
zone within an otherwise stable background stratification which, so the
original argument goes, leads to local overturning and ultimately a staircase. The
argument may be modified for the present experiments by saying that $|\Lambda|$
becomes locally so small in some layer that large scale convection is preferred over fingers
instead of expecting a convection roll as soon as $|\Lambda| < 1$
in some layer. The flux ratio induced instability may therefore be a serious
contender for explaining the staircase formation in the present experiments.

This hypothesis cannot be directly checked because the heat flux was not
measured. In order to nonetheless obtain some information about the flux ratio,
we will consider the situation prior to staircase formation in which fingers
span the entire layer and in which the fingers obey the scalings reported in
ref. \onlinecite{Hage10}. A theoretical consideration allows us to narrow down the
possible behaviour of the flux ratio in such a finger layer.

Double diffusive convection is described by the following set of equations for
the fields of velocity $\bm v(\bm r,t)$, concentration $c(\bm r,t)$, temperature
$T(\bm r,t)$ and pressure $p(\bm r,t)$:

\begin{equation}
\frac{\partial}{\partial t} \bm v + (\bm v \cdot \nabla) \bm v 
= - \frac{1}{\rho} \nabla p + \nu \nabla^2 \bm v + g \alpha T \hat {\bm z} 
-g \beta c \hat {\bm z} 
\label{eq:NSE}
\end{equation}
\begin{equation}
\nabla \cdot \bm v = 0
\end{equation}
\begin{equation}
\frac{\partial}{\partial t} T + \bm v \cdot \nabla T = \kappa \nabla^2 T 
\label{eq:T}
\end{equation}
\begin{equation}
\frac{\partial}{\partial t} c + \bm v \cdot \nabla c = D \nabla^2 c 
\end{equation}
where $\hat {\bm z}$ denotes a unit vector in the vertical direction and the
other variables are defined in section \ref{Experiment}. If we denote the
average over space and time by angular brackets, we find by taking the scalar
product of eq. (\ref{eq:NSE}) with $\bm v$ and subsequent averaging over space and
time that
\begin{equation}
\epsilon = - g \beta \langle v_z c \rangle + g \alpha \langle v_z T \rangle
\label{eq:energy}
\end{equation}
where $\epsilon$ is the dissipation defined as 
$\epsilon = \nu \sum_{i,j} \langle \left ( \frac{\partial v_j}{\partial x_i}
\right )^2 \rangle$.
If the entire layer is filled with uniform fingers, we may express the
definition of the flux ratio $\gamma$ within the finger in terms of the above
averages as
\begin{equation}
\gamma = \frac{\alpha \langle v_z T \rangle}{\beta \langle v_z c \rangle}.
\end{equation}
With the help of the previous equation, $\gamma$ is given by
\begin{equation}
\gamma =  1 +\frac{\epsilon}{g \beta \langle v_z c \rangle}.
\end{equation}
The denominator of the fraction is directly related to the measured Sherwood
number by
\begin{equation}
\langle v_z c \rangle = - D \frac{\Delta c}{L} (Sh-1).
\end{equation}
The dissipation $\epsilon$ on the other hand is not measured directly. However,
the flow within fingers is essentially a laminar parallel shear flow so that the
velocity profile is determined only by buoyancy and diffusion. For fingers of
thickness $d$, the dissipation is given by
$\epsilon = \nu K \langle v_z^2 \rangle / d^2$
with some geometric factor $K$ which depends on the velocity profile within the
fingers. For lamellar fingers with a sinusoidal profile and 
$v_z \propto \sin \left ( \frac{\pi}{d} x \right )$, one finds $K=\pi^2$.
For fingers of square cross section and
$v_z \propto \sin \left ( \frac{\pi}{d} x \right ) \sin \left ( \frac{\pi}{d} y \right )$
the factor $K$ is $K=2\pi^2$. The factor $K$ is larger if the velocity profile
includes higher harmonics. In summary, one expects $K$ to be a modest multiple
of $\pi^2$. Inserting the previous expressions into the formula for $\gamma$
leads to
\begin{equation}
\gamma =  1 - K \frac{Sc^2}{Ra_c} \frac{Re^2}{\left ( \frac{d}{L} \right )^2
(Sh-1)}
\end{equation}
with the approximation $\langle v_z^2 \rangle \approx \langle |\bm v|^2 \rangle$
which is reasonable for finger convection. We next wish to obtain an expression
for $\gamma$ in terms of the control parameters and have to eliminate the
experimental observables $Re$, $d/L$ and $Sh$ in favor of $Ra_c$ and $Ra_T$.
We will do so by invoking the scaling laws determined for pure finger convection
\cite{Hage10}:
\begin{equation}
\frac{d}{L} = 0.95 |Ra_T|^{-1/3} Ra_c^{1/9}
\label{eq:d_Ra}
\end{equation}
\begin{equation}
Re=10^{-6} |Ra_T|^{-1/2} Ra_c
\label{eq:Re_Ra}
\end{equation}
\begin{equation}
Sh=0.016 |Ra_T|^{-1/12} Ra_c^{4/9}
\label{eq:Sh_Ra}
\end{equation}
A theory put forward in ref. \onlinecite{Yang16} yields similar relationships, but as
this theory also depends on empirically adjusted parameters, we prefer the
scalings obtained from direct fits to the data. Inserting these into the
expression for $\gamma$ finally yields
\begin{equation}
\gamma = 1 - 69.25 \times 10^{-12} Sc^2 K |Ra_T|^{-1/4} Ra_c^{1/3}.
\label{eq:gamma1}
\end{equation}

One can deduce from this equation that the scalings 
(\ref{eq:d_Ra}-\ref{eq:Sh_Ra})
cannot be valid for all $Ra_c$ and $|Ra_T|$ because $\gamma>0$ must hold. To see
this, define the temperature deviation from the conduction profile $\theta$ as
$\theta = T - T_{\rm{bottom}} + \Delta T \frac{z}{L}$. The equation of evolution for
$\theta$ implied by eq. (\ref{eq:T}) is
\begin{equation}
\frac{\partial}{\partial t} \theta + \bm v \cdot \nabla \theta 
- \frac{\Delta T}{L} v_z = \kappa \nabla^2 \theta 
\end{equation}
and $\theta$ must be zero on the boundaries. Multiplying this equation by
$\theta$ and averaging over space and time yields
\begin{equation}
\frac{\Delta T}{L} \langle v_z \theta \rangle = \kappa \langle |\nabla \theta|^2
\rangle.
\end{equation}
Since $\langle v_z \theta \rangle = \langle v_z T \rangle$ and $\Delta T<0$, it
follows that $\langle v_z T \rangle<0$. The analogous sequence of operations
applied to the equation for $c$ yields $\langle v_z c \rangle<0$. As a
consequence, $\gamma>0$.

The condition $\gamma>0$ constrains $Ra_c$ and $|Ra_T|$ in eq. (\ref{eq:gamma1})
to
\begin{equation}
|Ra_T| \geq (69.25 \times 10^{-12} Sc^2 K)^4 Ra_c^{4/3}
\end{equation}
which is fulfilled for any reasonable choice of $K$ on the transition line
between fingers and staircases in fig. \ref{fig:param_space}. This means that
the condition $\gamma>0$ does not determine when fingers cease to exist and
staircases start to appear. We can therefore apply eq. (\ref{eq:gamma1}) at the
transition and transform it to
\begin{equation}
\gamma = 1 - 69.25 \times 10^{-12} Sc^2 \left( \frac{Sc}{Pr} \right)^{1/4} K
|\Lambda|^{-1/4} Ra_c^{1/12}.
\label{eq:gamma2}
\end{equation}
It is frequently assumed that $\gamma$ is a function of $|\Lambda|$ alone and
not of $Ra_c$ and $|Ra_T|$ independently. It is remarkable that, apart from a weak dependence
on $Ra_c$ in $Ra_c^{1/12}$, eq.
(\ref{eq:gamma2}) nearly fulfills this property noting that it was obtained from
the scalings eqs. (\ref{eq:d_Ra}-\ref{eq:Sh_Ra}) which themselves are the result of
three independent fits to experimental data. The data from which
(\ref{eq:d_Ra}-\ref{eq:Sh_Ra}) are deduced were obtained for approximately
constant $Sc$ and $Pr$ so that the dependence on $Sc$ and $Pr$ in eq.
(\ref{eq:gamma2}) is meaningless. On the other hand, eq. (\ref{eq:gamma2}) shows
that $\gamma$ is a monotonically increasing function of $|\Lambda|$ whenever
eqs. (\ref{eq:d_Ra}-\ref{eq:Sh_Ra}) are valid. There is therefore no indication of a
minimum in the flux ratio as a function of $|\Lambda|$ and the flux ratio
induced instability cannot be invoked as a mechanism for the staircase formation
in the experiments reported here.

Having excluded all standard explanations for the existence of staircases, the
question arises as to what actually caused them. A shear flow instability is not
plausible either. The Reynolds number based on the finger thickness 
$Re_d=Re (d/L)$ is too small to generate an instability. $Re_d$ is less than 1.8
in all experiments. In the so called Kolmogorov flow, which is an infinitely
extended parallel shear flow with a sinusoidal velocity profile, $Re_d$ would
have to be at least $\sqrt{2}\pi$ to cause instability \cite{Green74}.

\begin{figure}
\includegraphics[width=10cm]{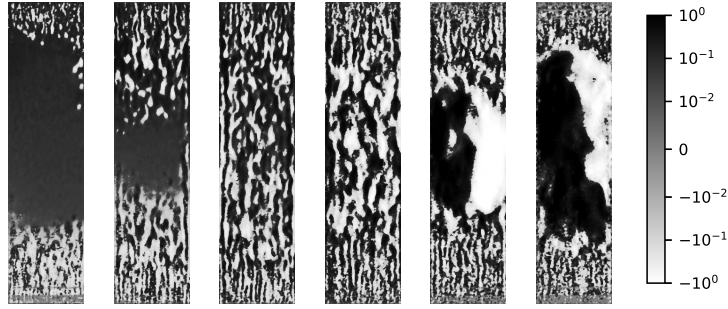}
\caption{Time evolution in a cell of $40cm$ height at $Ra_c=1.13 \times 10^{13}$
and $|Ra_T|=1.14 \times 10^{10}$ at times (from left to right) $t=8min$,
$14min$, $22min$, $115 min$, $216 min$ and $311 min$ after the voltage was
applied to the cell, and also after PIV particles were injected at the bottom.
These particles rise and lead to a mean upward velocity at early times.}
\label{fig:evolution}
\end{figure}

Another scenario could be that at the combinations of $Ra_c$ and $Ra_T$ of staircases,
the fingers never grow to full height if the voltage is switched on in the
experiments while the temperature gradient is already established. A behavior of
this type is observed in numerical simulations \cite{Singh14} of fingers
growing from an interface separating two homogeneous layers. In that case,
fingers grow only to a certain finite length. A hint at the true instability
mechanism may therefore be provided by how the flow evolves in time from initial
conditions to finally form a staircase. The temporal evolution recorded in
detail for $Ra_c=1.13 \times 10^{13}$ and $|Ra_T|=1.14 \times 10^{10}$ 
is shown in fig. \ref{fig:evolution}.
Starting from a uniform ion concentration and a linear temperature profile,
fingers appeared next to the electrodes as the voltage was applied. Two fronts
separating finger convection near each electrode from quiescent fluid in the
center propagated away from the boundaries
until they met in the middle of the cell. There was no mechanism at work to
limit finger growth. The fingers then progressively thickened in a central layer
until a single up and down flow were left over. Finally, the central layer
increased somewhat to further reduce the height of the fingers in the bordering
layers.

The increase of finger thickness reminds of the clustering of fingers which was 
found in numerical simulations \cite{Papare12} which reproduced staircases in the sense
of layers with alternating large and small gradients of temperature and salinity, 
but which did not show any layers with narrow fingers next to layers with broad structures.
The explanation for these staircases again relies on a stable density stratification 
and is not easily transferred to the present problem.

The sequence of events in fig. \ref{fig:evolution} suggests that layering may be
triggered by a random reduction of the stabilizing temperature gradient at some
height. Assuming eq. \ref{eq:d_Ra} also holds locally in the form
\begin{equation}
d= 0.95 \left(\frac{g \alpha}{\kappa \nu} \right)^{-1/3} 
\left(\frac{g \beta}{D \nu} \right)^{1/9}
\left(\frac{\Delta T}{L} \right)^{-1/3} \Delta c^{1/9}
,
\end{equation}
a reduced temperature gradient leads to wider fingers at that
height, which because of reduced
friction increases the flow velocity inside the fingers, which in turn enhances
the mixing and further reduces locally the temperature gradient to close a
positive feedback loop. The local reduction of the temperature gradient must
increase the temperature gradient in layers above and below the height at which
the instability started, which suppresses transport in the upper and lower
layers and saturates the instability. The example at $Ra_c=1.12 \times 10^{13}$ and
$|Ra_T|=4.97 \times 10^{10}$, in fig. \ref{fig:Ra_sequence}
would then simply show a case in which the unstable mode was prevented from
growing to an amplitude at which the stabilizing temperature gradient at the center is
reduced to the point that there is a single up and down flow in the central
layer of the staircase. This scenario superficially fits the evolution of
fingers in fig. \ref{fig:evolution}, but future studies will have to determine
whether this is a viable mechanism for staircase formation.

\begin{figure}
\includegraphics[width=10cm]{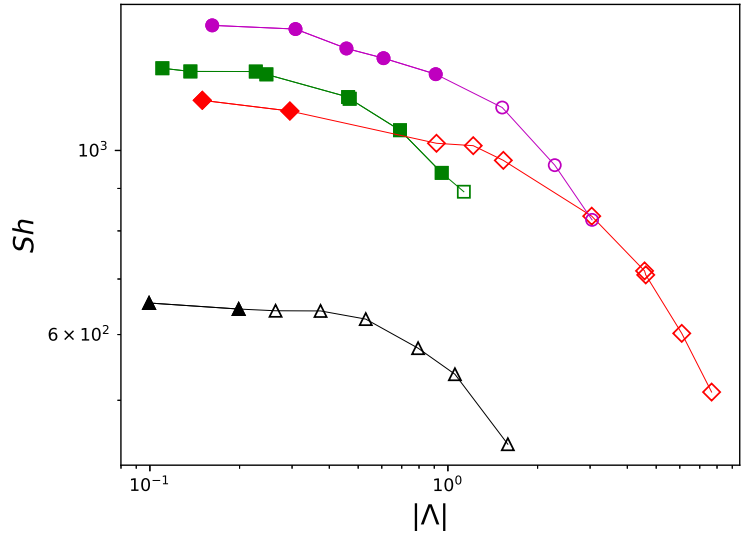}
\caption{(Color online) $Sh$ as a function of $|\Lambda|$ for $Ra_c=1.06 \times
10^{12}$ (triangles), $2.86 \times 10^{12}$ (diamonds), $5.71 \times 10^{12}$
(circles) and $2.04 \times 10^{13}$ (squares). Filled and open symbols represent staircases and
fingers crossing the whole cell, respectively.}
\label{fig:Knick}
\end{figure}

\section{Transport}
\label{Transport}

\subsection{Transport through the cell}

\begin{figure}
\includegraphics[width=10cm]{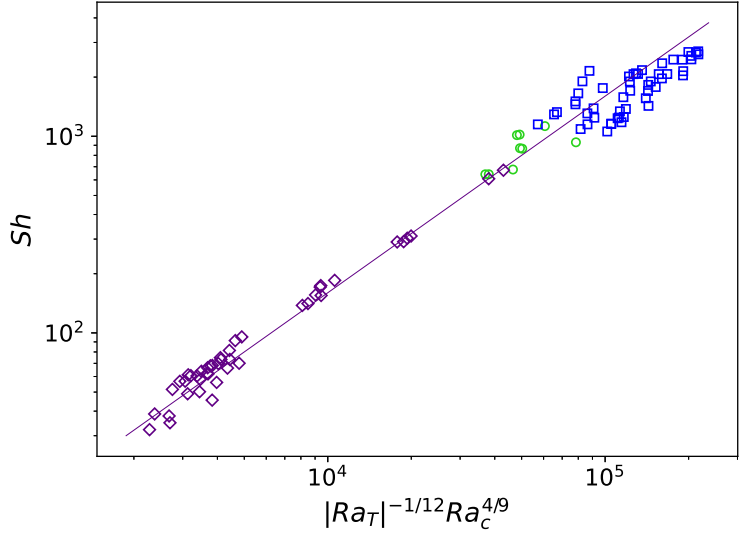}
\caption{(Color online) The Sherwood number $Sh$ as a function of $|Ra_T|^{-1/12} Ra_c^{4/9}$
for all data from ref. \onlinecite{Hage10} (diamonds), and the fingers (circles)
and staircases (squares) for which $Sh$ depends little on $|Ra_T|$ in the light
shaded area in fig. \ref{fig:param_space}. The solid line indicates 
$Sh=0.016 |Ra_T|^{-1/12} Ra_c^{4/9}$.}
\label{fig:Sh_global}
\end{figure}

The previous section showed that the scalings (\ref{eq:d_Ra}-\ref{eq:Sh_Ra})
cannot hold at all $Ra_c$ and $Ra_T$ because eq. (\ref{eq:gamma2}) is
incompatible with the requirement $\gamma>0$ at large $Ra_c$. Another limit
should
exist for the validity of eq. (\ref{eq:Sh_Ra}) at large $|Ra_T|$. For
$|\Lambda|=Pr/Sc$, or equivalently $|Ra_T|=Ra_c$, the system is linearly stable
and in absence of any instability to finite amplitude perturbations, all motion
must come to a stop and $Sh=1$. Eq. (\ref{eq:Sh_Ra}) however evaluates to
$Sh=0.016 Ra_c^{13/36}$ for $|Ra_T|=Ra_c$ which is far from 1. One therefore
expects eq. (\ref{eq:Sh_Ra}) to lose validity if $|Ra_T|$ is large enough to be
comparable with $Ra_c$. The different series of measurements carried out at
constant $Ra_c$ by varying $|Ra_T|$ recognizable in fig. \ref{fig:param_space}
probe the $|Ra_T|$ dependence of $Sh$. Fig. \ref{fig:Knick} shows $Sh$ for a few
$Ra_c$ as a function of $|\Lambda|$. The parameter $\Lambda$ 
is proportional to $Ra_T$ if $Ra_c$,
$Sc$ and $Pr$ are fixed. $Sh$ is weakly dependent on $|Ra_T|$ as
predicted by eq. (\ref{eq:Sh_Ra}) for small $|\Lambda|$, but if $|\Lambda|$ is
large enough, $Sh$ varies as expected more rapidly as a function of $|Ra_T|$. The full
function $Sh(Ra_c,|\Lambda|)$ is too complicated to be approximated by a simple
analytical expression because the transition from slow to fast variation as a
function of $|Ra_T|$ occurs at a different $|\Lambda|$ for different $Ra_c$. 
At any fixed $Ra_c$, the
tangents to $Sh(|\Lambda|)$ at small and large $|\Lambda|$ intersect at some
$|\Lambda|$. This intersection point marks the $|\Lambda|$ beyond which eq.
(\ref{eq:Sh_Ra}) is certainly not valid any more. The question is
whether it is always valid below that $|\Lambda|$. The
region of parameter space in which $Sh$ varies slowly as a function of $|Ra_T|$
is delimited in fig. \ref{fig:param_space}. Fig. \ref{fig:Sh_global} 
blindly applies eq. (\ref{eq:Sh_Ra}) to all measurements taken within this
region irrespective of whether the flow is a single finger layer or a
staircase. Within the scatter of the data, the Sherwood number for staircases
does not allow one to tell staircases from fingers crossing the entire layer.

However, fig. \ref{fig:Knick} shows that eq. (\ref{eq:Sh_Ra}) cannot be the
correct law for the Rayleigh number dependence of $Sh$ for the staircases. This
follows from the fact that for example near $|\Lambda|=0.2$, the dependence of
$Sh$ on $Ra_c$ is not monotonous in fig. \ref{fig:Knick}. Fig. \ref{fig:Knick}
also shows that there is no discontinuity in $Sh$ as the flow switches from a
single finger layer to a staircase. If this transition occurs within the region
in which $Sh$ depends little on $|Ra_T|$, as is the case for $Ra_c=1.06 \times
10^{12}$ in fig. \ref{fig:Knick}, one finds nearly identical $Sh$ for staircases
and pure finger flows. This example supports the conclusion that a measurement
of $Sh$ alone cannot distinguish fingers from staircases. This finding seems to
contradict the statement in ref. \onlinecite{Krishn03} that the Sherwood number
depends inversely proportional to the number $n$ of finger layers. However, the
behavior in $1/n$ is not seen for $n=1$ and $n=2$ in the data of ref.
\onlinecite{Krishn03} either, which contains data up to $n=8$, all obtained for
$|\Lambda|>1$.

\subsection{Transport through finger layers}

\begin{figure}
\includegraphics[width=10cm]{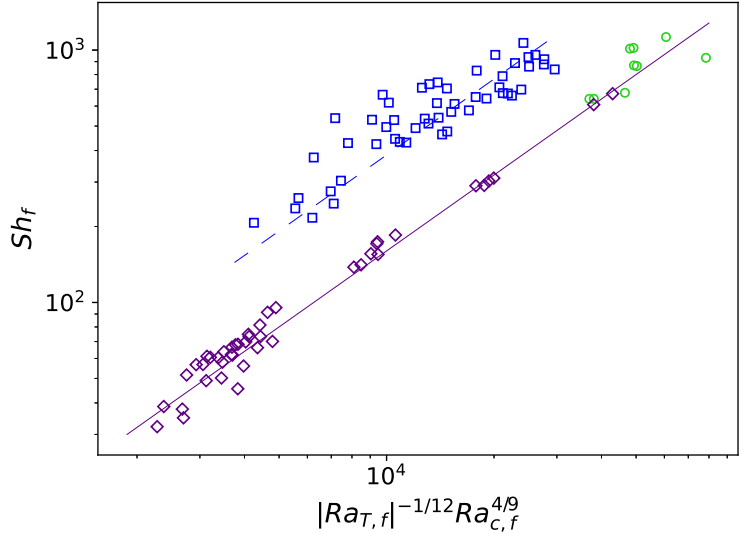}
\caption{(Color online) The finger Sherwood number $Sh_f$ as a function of $|Ra_{T,f}|^{-1/12} Ra_{c,f}^{4/9}$
for all data from ref. \onlinecite{Hage10} (diamonds), and the fingers (circles)
and staircases (squares) for which $Sh$ depends little on $|Ra_T|$ in the light
shaded area in fig. \ref{fig:param_space}. The solid line indicates
$Sh_f=0.016 |Ra_{T,f}|^{-1/12} Ra_{c,f}^{4/9}$, whereas the dashed line shows a power
law with the same exponents but with a prefactor fitted to the staircases
only, $Sh_f=0.0384 |Ra_{T,f}|^{-1/12} Ra_{c,f}^{4/9}$.}
\label{fig:Sh_finger}
\end{figure}

The goal of this section is to verify whether the ion transport through fingers
in a staircase is the same as through fingers which extend from top to bottom
boundaries, or whether a relation similar to eq. (\ref{eq:Sh_Ra}) also holds for
fingers in staircases. For that purpose, one has to define chemical and thermal
Rayleigh numbers $Ra_{T,f}$ and $Ra_{c,f}$ computed from the length of the
fingers $L_f$ and the difference in ion concentration $\Delta c_f$ and
temperature $\Delta T_f$ across the finger layer as
\begin{equation}
Ra_{T,f}=\frac{g \alpha \Delta T_f L_f^3}{\kappa \nu}  ~~~,~~~ 
Ra_{c,f}=\frac{g \beta \Delta c_f L_f^3}{D \nu}.
\end{equation}
PIV measurements yield $L_f$ (see fig. \ref{fig:vh_profile}) and the velocity
within the central layer. The Peclet number based on this velocity, the height
of the central layer and the ion diffusivity is in the range $10^4-10^6$ so
that the central layer is expected to efficiently mix concentration and to
annihilate any average concentration gradient in its interior. The concentration
drop across the top and bottom finger layers must then be 
$\Delta c_f = \Delta c/2$ each. The analogous argument for temperature leading
to $\Delta T_f = \Delta T/2$ is less compelling because the Peclet number for
temperature is smaller. We will nonetheless assume $\Delta T_f = \Delta T/2$.
To alleviate the impact of this assumption, we again
restrict ourselves to the same data points as those retained in fig.
\ref{fig:Sh_global} for which the temperature drop can be expected to only have
a weak effect on the ion transport as predicted by eq. (\ref{eq:Sh_Ra}). 
Fig. \ref{fig:Sh_finger} shows for this set of points that the finger Sherwood
number $Sh_f$ defined as
\begin{equation}
Sh_f = \frac{j\,L_f}{z\,F\,D\,\Delta c_f}
\end{equation}
is consistently larger for fingers in staircases than the Sherwood number 
for fingers crossing the entire cell. Note that the definitions of $Sh_f$,
$Ra_{T,f}$ and $Ra_{c,f}$ reduce to their previously defined counter parts $Sh$, 
$Ra_{T}$ and $Ra_{c}$ if one sets $L_f=L$, $\Delta c_f=\Delta c$ and
$\Delta T_f=\Delta T$ as is appropriate for fingers in a single finger layer.
Apart from the prefactor, the scaling (\ref{eq:Sh_Ra}) is also a plausible
summary of the Sherwood number measurements for fingers in staircases. 
A difference in prefactors is actually expected because of
the difference in boundary conditions that apply at the tips of the different
types of fingers.

\section{Conclusion}

The experiments reported here extend the parameter range covered by previous
experiments with the same system. This allows us to pin down more precisely what
determines the limit between finger convection and a single large scale
convection layer. The insufficient diffusive transport of heat between
neighboring fingers causes fingers to be replaced by standard convection flow.

Fingers may also give way to a staircase. The ion transport through the entire
fluid layer varies continuously through the transition and the 
Sherwood does not allow one to distinguish a staircase from a single finger
layer within the uncertainties of the data presented here. Despite
the similarity in the global Sherwood number, the ion transport behaves clearly 
differently depending on whether ions are transported through fingers in a
staircase or through fingers bridging the entire cell. The dependence of the
Sherwood number on the Rayleigh numbers previously determined \cite{Hage10} for
the latter type of fingers showed weak dependence on the thermal Rayleigh
number. This scaling does not extrapolate properly to the parameters of linear
onset of double diffusive convection, assuming that there indeed is no motion in
the linearly stable system. The new experiments detected a deviation from the
known scaling for sufficiently large density ratio.

The mechanism responsible for the formation of staircases, however, remains
elusive. None of the previously proposed mechanisms provides us with a
satisfactory explanation for the appearance of staircases in the experiments. It
is known since ref. \onlinecite{Hage10} that finger convection occurs at
$|\Lambda|<1$ even though the unstable stratification could just as well support
global overturning. It then is not surprising any more to find staircases at
$|\Lambda|<1$ as well.  Field observations in the oceans have so far detected both 
salt fingers and staircases only for $|\Lambda|>1$.
It will have to be decided in the future whether this is due to
a lack of observations or whether there is a more fundamental difference
between the oceanic staircases and laboratory experiments.


\begin{thebibliography}{28}%
\makeatletter
\providecommand \@ifxundefined [1]{%
 \@ifx{#1\undefined}
}%
\providecommand \@ifnum [1]{%
 \ifnum #1\expandafter \@firstoftwo
 \else \expandafter \@secondoftwo
 \fi
}%
\providecommand \@ifx [1]{%
 \ifx #1\expandafter \@firstoftwo
 \else \expandafter \@secondoftwo
 \fi
}%
\providecommand \natexlab [1]{#1}%
\providecommand \enquote  [1]{``#1''}%
\providecommand \bibnamefont  [1]{#1}%
\providecommand \bibfnamefont [1]{#1}%
\providecommand \citenamefont [1]{#1}%
\providecommand \href@noop [0]{\@secondoftwo}%
\providecommand \href [0]{\begingroup \@sanitize@url \@href}%
\providecommand \@href[1]{\@@startlink{#1}\@@href}%
\providecommand \@@href[1]{\endgroup#1\@@endlink}%
\providecommand \@sanitize@url [0]{\catcode `\\12\catcode `\$12\catcode
  `\&12\catcode `\#12\catcode `\^12\catcode `\_12\catcode `\%12\relax}%
\providecommand \@@startlink[1]{}%
\providecommand \@@endlink[0]{}%
\providecommand \url  [0]{\begingroup\@sanitize@url \@url }%
\providecommand \@url [1]{\endgroup\@href {#1}{\urlprefix }}%
\providecommand \urlprefix  [0]{URL }%
\providecommand \Eprint [0]{\href }%
\providecommand \doibase [0]{http://dx.doi.org/}%
\providecommand \selectlanguage [0]{\@gobble}%
\providecommand \bibinfo  [0]{\@secondoftwo}%
\providecommand \bibfield  [0]{\@secondoftwo}%
\providecommand \translation [1]{[#1]}%
\providecommand \BibitemOpen [0]{}%
\providecommand \bibitemStop [0]{}%
\providecommand \bibitemNoStop [0]{.\EOS\space}%
\providecommand \EOS [0]{\spacefactor3000\relax}%
\providecommand \BibitemShut  [1]{\csname bibitem#1\endcsname}%
\let\auto@bib@innerbib\@empty
\bibitem [{\citenamefont {Radko}(2013)}]{Radko13}%
  \BibitemOpen
  \bibfield  {author} {\bibinfo {author} {\bibfnamefont {T.}~\bibnamefont
  {Radko}},\ }\href {\doibase 10.1017/CBO9781139034173} {\emph {\bibinfo
  {title} {Double-Diffusive Convection}}}\ (\bibinfo  {publisher} {Cambridge
  University Press},\ \bibinfo {year} {2013})\BibitemShut {NoStop}%
\bibitem [{\citenamefont {Krishnamurti}(2003)}]{Krishn03}%
  \BibitemOpen
  \bibfield  {author} {\bibinfo {author} {\bibfnamefont {R.}~\bibnamefont
  {Krishnamurti}},\ }\bibfield  {title} {\enquote {\bibinfo {title}
  {Double-diffusive transport in laboratory thermohaline staircases},}\
  }\href@noop {} {\bibfield  {journal} {\bibinfo  {journal} {J. Fluid Mech.}\
  }\textbf {\bibinfo {volume} {483}},\ \bibinfo {pages} {287--314} (\bibinfo
  {year} {2003})}\BibitemShut {NoStop}%
\bibitem [{\citenamefont {Krishnamurti}(2009)}]{Krishn09}%
  \BibitemOpen
  \bibfield  {author} {\bibinfo {author} {\bibfnamefont {R.}~\bibnamefont
  {Krishnamurti}},\ }\bibfield  {title} {\enquote {\bibinfo {title} {Heat, salt
  and momentum transfer in a laboratory thermohaline staircase},}\ }\href@noop
  {} {\bibfield  {journal} {\bibinfo  {journal} {J. Fluid Mech.}\ }\textbf
  {\bibinfo {volume} {638}},\ \bibinfo {pages} {491--506} (\bibinfo {year}
  {2009})}\BibitemShut {NoStop}%
\bibitem [{\citenamefont {Hage}\ and\ \citenamefont {Tilgner}(2010)}]{Hage10}%
  \BibitemOpen
  \bibfield  {author} {\bibinfo {author} {\bibfnamefont {E.}~\bibnamefont
  {Hage}}\ and\ \bibinfo {author} {\bibfnamefont {A.}~\bibnamefont {Tilgner}},\
  }\bibfield  {title} {\enquote {\bibinfo {title} {{High Rayleigh number
  convection with double diffusive fingers}},}\ }\href@noop {} {\bibfield
  {journal} {\bibinfo  {journal} {Phys. Fluids}\ }\textbf {\bibinfo {volume}
  {22}},\ \bibinfo {pages} {076603} (\bibinfo {year} {2010})}\BibitemShut
  {NoStop}%
\bibitem [{\citenamefont {Kellner}\ and\ \citenamefont
  {Tilgner}(2014)}]{Kellne14}%
  \BibitemOpen
  \bibfield  {author} {\bibinfo {author} {\bibfnamefont {M.}~\bibnamefont
  {Kellner}}\ and\ \bibinfo {author} {\bibfnamefont {A.}~\bibnamefont
  {Tilgner}},\ }\bibfield  {title} {\enquote {\bibinfo {title} {{Transition to
  finger convection in double-diffusive convection}},}\ }\href@noop {}
  {\bibfield  {journal} {\bibinfo  {journal} {Phys. Fluids}\ }\textbf {\bibinfo
  {volume} {26}},\ \bibinfo {pages} {094103} (\bibinfo {year}
  {2014})}\BibitemShut {NoStop}%
\bibitem [{\citenamefont {Goldstein}, \citenamefont {Chiang},\ and\
  \citenamefont {See}(1990)}]{Goldst90}%
  \BibitemOpen
  \bibfield  {author} {\bibinfo {author} {\bibfnamefont {R.~J.}\ \bibnamefont
  {Goldstein}}, \bibinfo {author} {\bibfnamefont {H.~D.}\ \bibnamefont
  {Chiang}}, \ and\ \bibinfo {author} {\bibfnamefont {D.~L.}\ \bibnamefont
  {See}},\ }\bibfield  {title} {\enquote {\bibinfo {title}
  {{High-Rayleigh-number convection in a horizontal enclosure}},}\ }\href@noop
  {} {\bibfield  {journal} {\bibinfo  {journal} {J. Fluid Mech.}\ }\textbf
  {\bibinfo {volume} {213}},\ \bibinfo {pages} {111--126} (\bibinfo {year}
  {1990})}\BibitemShut {NoStop}%
\bibitem [{\citenamefont {Kumar}, \citenamefont {Srivastava},\ and\
  \citenamefont {Karagadde}(2018)}]{Kumar18}%
  \BibitemOpen
  \bibfield  {author} {\bibinfo {author} {\bibfnamefont {V.}~\bibnamefont
  {Kumar}}, \bibinfo {author} {\bibfnamefont {A.}~\bibnamefont {Srivastava}}, \
  and\ \bibinfo {author} {\bibfnamefont {S.}~\bibnamefont {Karagadde}},\
  }\bibfield  {title} {\enquote {\bibinfo {title} {Compositional dependency of
  double-diffusive layers during binary alloy solidification: Full-field
  measurements and quantification},}\ }\href {\doibase 10.1063/1.5049135}
  {\bibfield  {journal} {\bibinfo  {journal} {Physics of Fluids}\ }\textbf
  {\bibinfo {volume} {30}},\ \bibinfo {pages} {113603} (\bibinfo {year}
  {2018})}\BibitemShut {NoStop}%
\bibitem [{\citenamefont {Radko}(2020)}]{Radko20}%
  \BibitemOpen
  \bibfield  {author} {\bibinfo {author} {\bibfnamefont {T.}~\bibnamefont
  {Radko}},\ }\bibfield  {title} {\enquote {\bibinfo {title} {Suppression of
  internal waves by thermohaline staircases},}\ }\href {\doibase
  10.1017/jfm.2020.563} {\bibfield  {journal} {\bibinfo  {journal} {Journal of
  Fluid Mechanics}\ }\textbf {\bibinfo {volume} {902}},\ \bibinfo {pages} {A14}
  (\bibinfo {year} {2020})}\BibitemShut {NoStop}%
\bibitem [{\citenamefont {Radko}(2016)}]{Radko16}%
  \BibitemOpen
  \bibfield  {author} {\bibinfo {author} {\bibfnamefont {T.}~\bibnamefont
  {Radko}},\ }\bibfield  {title} {\enquote {\bibinfo {title} {Thermohaline
  layering in dynamically and diffusively stable shear flows},}\ }\href
  {\doibase 10.1017/jfm.2016.547} {\bibfield  {journal} {\bibinfo  {journal}
  {Journal of Fluid Mechanics}\ }\textbf {\bibinfo {volume} {805}},\ \bibinfo
  {pages} {147–170} (\bibinfo {year} {2016})}\BibitemShut {NoStop}%
\bibitem [{\citenamefont {Brown}\ and\ \citenamefont {Radko}(2019)}]{Brown19}%
  \BibitemOpen
  \bibfield  {author} {\bibinfo {author} {\bibfnamefont {J.~M.}\ \bibnamefont
  {Brown}}\ and\ \bibinfo {author} {\bibfnamefont {T.}~\bibnamefont {Radko}},\
  }\bibfield  {title} {\enquote {\bibinfo {title} {Initiation of diffusive
  layering by time-dependent shear},}\ }\href {\doibase 10.1017/jfm.2018.790}
  {\bibfield  {journal} {\bibinfo  {journal} {Journal of Fluid Mechanics}\
  }\textbf {\bibinfo {volume} {858}},\ \bibinfo {pages} {588–608} (\bibinfo
  {year} {2019})}\BibitemShut {NoStop}%
\bibitem [{\citenamefont {Yang}\ \emph {et~al.}(2022)\citenamefont {Yang},
  \citenamefont {Verzicco}, \citenamefont {Lohse},\ and\ \citenamefont
  {Caulfield}}]{Yang22}%
  \BibitemOpen
  \bibfield  {author} {\bibinfo {author} {\bibfnamefont {Y.}~\bibnamefont
  {Yang}}, \bibinfo {author} {\bibfnamefont {R.}~\bibnamefont {Verzicco}},
  \bibinfo {author} {\bibfnamefont {D.}~\bibnamefont {Lohse}}, \ and\ \bibinfo
  {author} {\bibfnamefont {C.}~\bibnamefont {Caulfield}},\ }\bibfield  {title}
  {\enquote {\bibinfo {title} {Layering and vertical transport in sheared
  double-diffusive convection in the diffusive regime},}\ }\href {\doibase
  10.1017/jfm.2021.1091} {\bibfield  {journal} {\bibinfo  {journal} {Journal of
  Fluid Mechanics}\ }\textbf {\bibinfo {volume} {933}},\ \bibinfo {pages} {A30}
  (\bibinfo {year} {2022})}\BibitemShut {NoStop}%
\bibitem [{\citenamefont {Ma}\ and\ \citenamefont {Peltier}(2022)}]{Ma22}%
  \BibitemOpen
  \bibfield  {author} {\bibinfo {author} {\bibfnamefont {Y.}~\bibnamefont
  {Ma}}\ and\ \bibinfo {author} {\bibfnamefont {W.}~\bibnamefont {Peltier}},\
  }\bibfield  {title} {\enquote {\bibinfo {title} {Thermohaline staircase
  formation in the diffusive convection regime: a theory based upon stratified
  turbulence asymptotics},}\ }\href {\doibase 10.1017/jfm.2021.945} {\bibfield
  {journal} {\bibinfo  {journal} {Journal of Fluid Mechanics}\ }\textbf
  {\bibinfo {volume} {931}},\ \bibinfo {pages} {R4} (\bibinfo {year}
  {2022})}\BibitemShut {NoStop}%
\bibitem [{\citenamefont {Radko}(2003)}]{Radko03}%
  \BibitemOpen
  \bibfield  {author} {\bibinfo {author} {\bibfnamefont {T.}~\bibnamefont
  {Radko}},\ }\bibfield  {title} {\enquote {\bibinfo {title} {A mechanism for
  layer formation in a double-diffusive fluid},}\ }\href {\doibase
  10.1017/S0022112003006785} {\bibfield  {journal} {\bibinfo  {journal}
  {Journal of Fluid Mechanics}\ }\textbf {\bibinfo {volume} {497}},\ \bibinfo
  {pages} {365–380} (\bibinfo {year} {2003})}\BibitemShut {NoStop}%
\bibitem [{\citenamefont {Sellmach}\ \emph {et~al.}(2011)\citenamefont
  {Sellmach}, \citenamefont {Traxler}, \citenamefont {Garaud}, \citenamefont
  {Brummell},\ and\ \citenamefont {Radko}}]{Stellm11}%
  \BibitemOpen
  \bibfield  {author} {\bibinfo {author} {\bibfnamefont {S.}~\bibnamefont
  {Sellmach}}, \bibinfo {author} {\bibfnamefont {A.}~\bibnamefont {Traxler}},
  \bibinfo {author} {\bibfnamefont {P.}~\bibnamefont {Garaud}}, \bibinfo
  {author} {\bibfnamefont {N.}~\bibnamefont {Brummell}}, \ and\ \bibinfo
  {author} {\bibfnamefont {T.}~\bibnamefont {Radko}},\ }\bibfield  {title}
  {\enquote {\bibinfo {title} {Dynamics of fingering convection. part 2 the
  formation of thermohaline staircases},}\ }\href {\doibase
  10.1017/jfm.2011.99} {\bibfield  {journal} {\bibinfo  {journal} {Journal of
  Fluid Mechanics}\ }\textbf {\bibinfo {volume} {677}},\ \bibinfo {pages}
  {554–571} (\bibinfo {year} {2011})}\BibitemShut {NoStop}%
\bibitem [{\citenamefont {Traxler}\ \emph {et~al.}(2011)\citenamefont
  {Traxler}, \citenamefont {Stellmach}, \citenamefont {Garaud}, \citenamefont
  {Radko},\ and\ \citenamefont {Brummel}}]{Traxle11}%
  \BibitemOpen
  \bibfield  {author} {\bibinfo {author} {\bibfnamefont {A.}~\bibnamefont
  {Traxler}}, \bibinfo {author} {\bibfnamefont {S.}~\bibnamefont {Stellmach}},
  \bibinfo {author} {\bibfnamefont {P.}~\bibnamefont {Garaud}}, \bibinfo
  {author} {\bibfnamefont {T.}~\bibnamefont {Radko}}, \ and\ \bibinfo {author}
  {\bibfnamefont {N.}~\bibnamefont {Brummel}},\ }\bibfield  {title} {\enquote
  {\bibinfo {title} {Dynamics of fingering convection. part 1 small-scale
  fluxes and large-scale instabilities},}\ }\href {\doibase
  10.1017/jfm.2011.98} {\bibfield  {journal} {\bibinfo  {journal} {Journal of
  Fluid Mechanics}\ }\textbf {\bibinfo {volume} {677}},\ \bibinfo {pages}
  {530–553} (\bibinfo {year} {2011})}\BibitemShut {NoStop}%
\bibitem [{\citenamefont {Ma}\ and\ \citenamefont {Peltier}(2021)}]{Ma21}%
  \BibitemOpen
  \bibfield  {author} {\bibinfo {author} {\bibfnamefont {Y.}~\bibnamefont
  {Ma}}\ and\ \bibinfo {author} {\bibfnamefont {W.~R.}\ \bibnamefont
  {Peltier}},\ }\bibfield  {title} {\enquote {\bibinfo {title} {Gamma
  instability in an inhomogeneous environment and salt-fingering staircase
  trapping: Determining the step size},}\ }\href {\doibase
  10.1103/PhysRevFluids.6.033903} {\bibfield  {journal} {\bibinfo  {journal}
  {Phys. Rev. Fluids}\ }\textbf {\bibinfo {volume} {6}},\ \bibinfo {pages}
  {033903} (\bibinfo {year} {2021})}\BibitemShut {NoStop}%
\bibitem [{\citenamefont {Yang}\ \emph {et~al.}(2020)\citenamefont {Yang},
  \citenamefont {Chen}, \citenamefont {Verzicco},\ and\ \citenamefont
  {Lohse}}]{Yang20}%
  \BibitemOpen
  \bibfield  {author} {\bibinfo {author} {\bibfnamefont {Y.}~\bibnamefont
  {Yang}}, \bibinfo {author} {\bibfnamefont {W.}~\bibnamefont {Chen}}, \bibinfo
  {author} {\bibfnamefont {R.}~\bibnamefont {Verzicco}}, \ and\ \bibinfo
  {author} {\bibfnamefont {D.}~\bibnamefont {Lohse}},\ }\bibfield  {title}
  {\enquote {\bibinfo {title} {Multiple states and transport properties of
  double-diffusive convection turbulence},}\ }\href@noop {} {\bibfield
  {journal} {\bibinfo  {journal} {Proc. Nat. Acad. Sci.}\ }\textbf {\bibinfo
  {volume} {117}},\ \bibinfo {pages} {14676--14681} (\bibinfo {year}
  {2020})}\BibitemShut {NoStop}%
\bibitem [{\citenamefont {Radko}(2014)}]{Radko14}%
  \BibitemOpen
  \bibfield  {author} {\bibinfo {author} {\bibfnamefont {T.}~\bibnamefont
  {Radko}},\ }\bibfield  {title} {\enquote {\bibinfo {title} {Applicability and
  failure of the flux-gradient laws in double-diffusive convection},}\ }\href
  {\doibase 10.1017/jfm.2014.244} {\bibfield  {journal} {\bibinfo  {journal}
  {Journal of Fluid Mechanics}\ }\textbf {\bibinfo {volume} {750}},\ \bibinfo
  {pages} {33–72} (\bibinfo {year} {2014})}\BibitemShut {NoStop}%
\bibitem [{\citenamefont {Radko}(2019)}]{Radko19b}%
  \BibitemOpen
  \bibfield  {author} {\bibinfo {author} {\bibfnamefont {T.}~\bibnamefont
  {Radko}},\ }\bibfield  {title} {\enquote {\bibinfo {title} {Thermohaline
  layering on the microscale},}\ }\href {\doibase 10.1017/jfm.2018.976}
  {\bibfield  {journal} {\bibinfo  {journal} {Journal of Fluid Mechanics}\
  }\textbf {\bibinfo {volume} {862}},\ \bibinfo {pages} {672–695} (\bibinfo
  {year} {2019})}\BibitemShut {NoStop}%
\bibitem [{\citenamefont {Probstein}(1995)}]{Probst95}%
  \BibitemOpen
  \bibfield  {author} {\bibinfo {author} {\bibfnamefont {R.~F.}\ \bibnamefont
  {Probstein}},\ }\href@noop {} {\emph {\bibinfo {title} {{Physicochemical
  Hydrodynamics}}}}\ (\bibinfo  {publisher} {John Wiley \& Sons},\ \bibinfo
  {address} {New York},\ \bibinfo {year} {1995})\BibitemShut {NoStop}%
\bibitem [{\citenamefont {Merryfield}(2000)}]{Merryf00}%
  \BibitemOpen
  \bibfield  {author} {\bibinfo {author} {\bibfnamefont {W.~J.}\ \bibnamefont
  {Merryfield}},\ }\bibfield  {title} {\enquote {\bibinfo {title} {Origin of
  thermohaline staircases},}\ }\href@noop {} {\bibfield  {journal} {\bibinfo
  {journal} {Jornal of Physical Oceanography}\ }\textbf {\bibinfo {volume}
  {30}},\ \bibinfo {pages} {1046--1068} (\bibinfo {year} {2000})}\BibitemShut
  {NoStop}%
\bibitem [{\citenamefont {Stern}(1969)}]{Stern69}%
  \BibitemOpen
  \bibfield  {author} {\bibinfo {author} {\bibfnamefont {M.~E.}\ \bibnamefont
  {Stern}},\ }\bibfield  {title} {\enquote {\bibinfo {title} {Collective
  instability of salt fingers},}\ }\href@noop {} {\bibfield  {journal}
  {\bibinfo  {journal} {J. Fluid Mech.}\ }\textbf {\bibinfo {volume} {35}},\
  \bibinfo {pages} {209--218} (\bibinfo {year} {1969})}\BibitemShut {NoStop}%
\bibitem [{\citenamefont {Holyer}(1981)}]{Holyer81}%
  \BibitemOpen
  \bibfield  {author} {\bibinfo {author} {\bibfnamefont {J.~Y.}\ \bibnamefont
  {Holyer}},\ }\bibfield  {title} {\enquote {\bibinfo {title} {{On the
  collective instability of salt fingers}},}\ }\href@noop {} {\bibfield
  {journal} {\bibinfo  {journal} {J. Fluid Mech.}\ }\textbf {\bibinfo {volume}
  {110}} (\bibinfo {year} {1981})}\BibitemShut {NoStop}%
\bibitem [{\citenamefont {Kunze}(1987)}]{Kunze87}%
  \BibitemOpen
  \bibfield  {author} {\bibinfo {author} {\bibfnamefont {E.}~\bibnamefont
  {Kunze}},\ }\bibfield  {title} {\enquote {\bibinfo {title} {{Limits on
  growing, finite-length salt fingers: A Richardson number constraint}},}\
  }\href@noop {} {\bibfield  {journal} {\bibinfo  {journal} {Journal of Marine
  Research}\ }\textbf {\bibinfo {volume} {45}},\ \bibinfo {pages} {533--556}
  (\bibinfo {year} {1987})}\BibitemShut {NoStop}%
\bibitem [{\citenamefont {Yang}, \citenamefont {Verzicco},\ and\ \citenamefont
  {Lohse}(2016)}]{Yang16}%
  \BibitemOpen
  \bibfield  {author} {\bibinfo {author} {\bibfnamefont {Y.}~\bibnamefont
  {Yang}}, \bibinfo {author} {\bibfnamefont {R.}~\bibnamefont {Verzicco}}, \
  and\ \bibinfo {author} {\bibfnamefont {D.}~\bibnamefont {Lohse}},\ }\bibfield
   {title} {\enquote {\bibinfo {title} {Scaling laws and flow structures of
  double diffusive convection in the finger regime},}\ }\href {\doibase
  10.1017/jfm.2016.484} {\bibfield  {journal} {\bibinfo  {journal} {Journal of
  Fluid Mechanics}\ }\textbf {\bibinfo {volume} {802}},\ \bibinfo {pages}
  {667–689} (\bibinfo {year} {2016})}\BibitemShut {NoStop}%
\bibitem [{\citenamefont {Green}(1974)}]{Green74}%
  \BibitemOpen
  \bibfield  {author} {\bibinfo {author} {\bibfnamefont {J.~S.~A.}\
  \bibnamefont {Green}},\ }\bibfield  {title} {\enquote {\bibinfo {title}
  {Two-dimensional turbulence near the viscous limit},}\ }\href@noop {}
  {\bibfield  {journal} {\bibinfo  {journal} {J. Fluid Mech.}\ }\textbf
  {\bibinfo {volume} {62}},\ \bibinfo {pages} {273--287} (\bibinfo {year}
  {1974})}\BibitemShut {NoStop}%
\bibitem [{\citenamefont {Singh}\ and\ \citenamefont
  {Srinivasan}(2014)}]{Singh14}%
  \BibitemOpen
  \bibfield  {author} {\bibinfo {author} {\bibfnamefont {O.~P.}\ \bibnamefont
  {Singh}}\ and\ \bibinfo {author} {\bibfnamefont {J.}~\bibnamefont
  {Srinivasan}},\ }\bibfield  {title} {\enquote {\bibinfo {title} {{Effect of
  Rayleigh numbers on the evolution of double-diffusive salt fingers}},}\
  }\href {\doibase 10.1063/1.4882264} {\bibfield  {journal} {\bibinfo
  {journal} {Physics of Fluids}\ }\textbf {\bibinfo {volume} {26}},\ \bibinfo
  {pages} {062104} (\bibinfo {year} {2014})}\BibitemShut {NoStop}%
\bibitem [{\citenamefont {Paparella}\ and\ \citenamefont {von
  Hardenberg}(2012)}]{Papare12}%
  \BibitemOpen
  \bibfield  {author} {\bibinfo {author} {\bibfnamefont {F.}~\bibnamefont
  {Paparella}}\ and\ \bibinfo {author} {\bibfnamefont {J.}~\bibnamefont {von
  Hardenberg}},\ }\bibfield  {title} {\enquote {\bibinfo {title} {Clustering of
  salt fingers in double-diffusive convection leads to staircaselike
  stratification},}\ }\href {\doibase 10.1103/PhysRevLett.109.014502}
  {\bibfield  {journal} {\bibinfo  {journal} {Phys. Rev. Lett.}\ }\textbf
  {\bibinfo {volume} {109}},\ \bibinfo {pages} {014502} (\bibinfo {year}
  {2012})}\BibitemShut {NoStop}%
\end{thebibliography}

%

\end{document}